\documentclass{ws-procs9x6}
\usepackage{epsfig}

\begin{document}

\title{\hfill{\tiny FZJ--IKP(TH)--2004--17, HISKP-TH-04-21} \\[1.8em]
Remarks
  on the parity determination \\ of narrow 
resonances\footnote{\uppercase{T}his work
    is partly supported by \uppercase{COSY} grant No. 41445282}}

\author{\underline{C. Hanhart}$^1$, J. Haidenbauer$^1$, K. Nakayama$^{1,2}$,
  U.-G.  Mei{\ss}ner$^{1,3}$}

\address{ $^1$Institut f\"{u}r Kernphysik, Forschungszentrum J\"{u}lich GmbH,\\ 
D--52425 J\"{u}lich, Germany \\
$^2$ Dept. of Physics and Astronomy, University of Georgia,\\ 
Athens, Georgia 30602, USA\\ 
$^3$Helmholtz-Institut f\"{u}r Strahlen- und Kernphysik (Theorie), 
Universit\"at Bonn \\ 
Nu{\ss}allee 14-16, D--53115 Bonn, Germany
}

\maketitle

\abstracts{Recently several proposals were put forward to determine the parity
  of narrow baryonic resonances, in particular the $\Theta^+$. In these
  proceedings we will briefly comment on the general problems in this task and
  then discuss in detail the potential of reactions of the type $\vec N\vec
  N\to\Theta^+Y$, where $Y$ denotes a hyperon (either $\Sigma$ or $\Lambda$)
  and the arrows indicate that a polarized initial state is required.
  Besides reiterating the model--independent properties of this class of
  reactions we discuss the physics content of some model calculations.}

\section{Generalities}

The parity of a hadron contains significant 
information on its substructure. Unfortunately, especially for spin--1/2
particles, the determination of the parity is a non--trivial problem
especially when we talk about narrow states.

To illustrate the origin of the difficulty we observe that---in leading order
of the outgoing cms momentum---the decay vertex of a spin--1/2 resonance into
a spin--1/2 particle and a pseudoscalar (e.g. $\Theta^+\to NK$, if the
$\Theta^+$ were indeed a spin--1/2 particle as suggested by almost all models)
reads $\vec \sigma \cdot \vec q$ for a positive parity resonance and $1$ for a
negative parity resonance. Thus, as long as we do not measure the polarization
of the decay products, all observables scale as $(\vec \sigma \cdot \vec q)
(\vec \sigma \cdot \vec q)=\vec q\cdot \vec q$ or $1$, accordingly, leaving no
unique trace of the intrinsic parity of the decaying object.

Unfortunately a measurement of the polarization of a nucleon in the final
state is technically very demanding and of low efficiency. In addition, as was
clearly demonstrated by Titov in this conference\cite{titov}, possible
interference phenomena with the background amplitudes put into question, if
such a measurement would indeed allow to determine the parity of the resonance
unambiguously.

Thus, the only straightforward option that remains is to extract the quantum
numbers of this narrow resonance from a partial wave analysis. 
However, if the
resonance of interest is very narrow also this program might be unsuccessful:
in this case the partial wave analysis allows solely to put an upper limit on
the width of the decaying particle\cite{arndt,haidenbauer}---note, also the
more sophisticated work of Ref.\cite{sib}, where data on $K^+d$ scattering is
used directly--- here the $KN$ partial wave analysis was used to fix the
parameters of the model for the background amplitudes.

Does this mean that there is no way to determine the parity of a narrow
resonance? The answer to this question is no, for one can use the stringent
selection rules that are enforced by the Pauli Principle on the
nucleon--nucleon ($NN$) systems to manipulate the total parity of a system. It
is well known that a two--nucleon state acquires a phase $(-)^{L+S+T}$ under
permutation of the two particles, where $L$, $S$, and $T$ denote the angular
momentum, the total spin and the total isospin of the two nucleon system. The
required antisymmetry of the $NN$ wavefunction thus calls for $L+S+T$ to be
odd. For example, for a proton--proton state $T=1$ and the parity is given by
$(-)^L$---thus each $S=1$ state has odd parity and each $S=0$ state has even
parity. Therefore, preparing a pure spin state of a $pp$ system means
preparing a $NN$ state of  known parity. In case of a $T=0$ state, the
assignment of spin and parity needs to be reversed.

A well known textbook example that exploits this method is the measurement of
the parity of the pion\cite{pion} in $\pi^-$ capture on deuterium from an
atomic $s$--state ($\pi^-d\to nn$); this transition would be forbidden if the
parity of the spin 0 pion were positive.  To see this we recall that an even
parity $nn$ state, characterized by an even value of $L$, is to be a spin
singlet; consequently, in this case we have for the total angular momentum
$J=L$. On the other hand, since for the deuteron $j=1$ the initial state has
$J=1$. Since the parity of the whole system agrees to that of the pion,
an $s$--wave production is allowed only for negative parity
pions.

It was essential in case of the reaction $d\pi^-\to nn$ that the deuteron is a
$j=1$ state and the pion is a scalar---this fixed the total angular momentum
to $J=1$ for the $s$--wave initial state. For the system $NN\to Y\Theta^+$ the
final state can be both spin singlet as well as spin triplet. Therefore in this
case we need to manipulate the spin of the initial state to fix its parity.
  This observation was first
made in Ref.\cite{thomas} and then further exploited in a series of
publications\cite{MODELS1,unsers,nam,nam2,uzikov1,uzikov2,unser2}.
In addition, we will have to use the energy dependence of the spin cross
sections to identify the leading partial wave, as was observed in 
Refs.\cite{thomas,unser2}.

\begin{figure}[t!]
\begin{center}
\epsfig{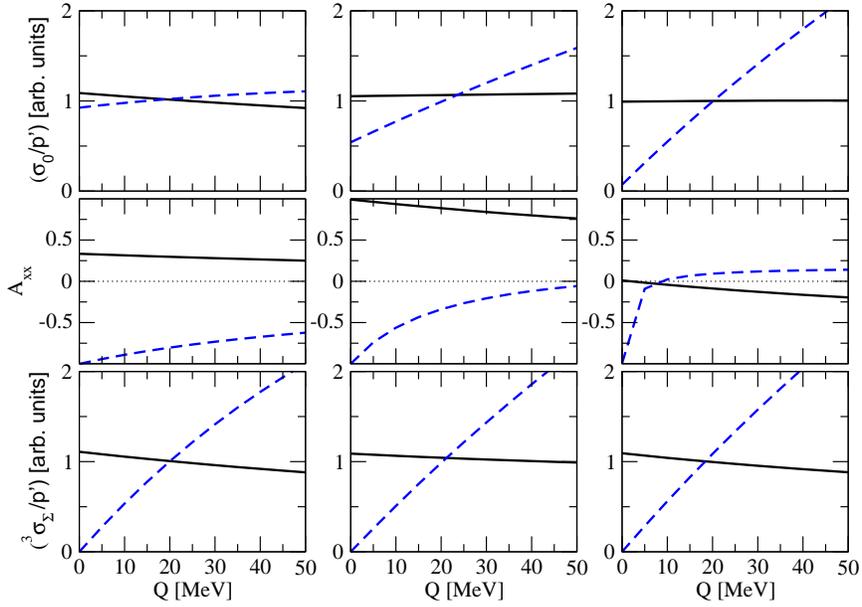}
\caption{\it Energy dependence of the total cross section and the angular
integrated polarization observable $A_{xx}$ and $^3\sigma_\Sigma$
for the reaction $pp\to \Sigma^+\Theta^+$.
Solid (dashed)
lines correspond to a negative (positive) parity $\Theta^+$. Shown are
results for 
three different models for the production operator: the left column shows the
results for the model with only kaon exchange, the middle one those for the
one with $K^*$ exchange and the right one those for the model including $K^*$ 
and $K$ exchange. All results for $\sigma_0$ and $^3\sigma_\Sigma$ are 
normalized to 1 at an excess energy of 20 MeV and are
divided by the phase-space volume. 
}
\label{int}
\end{center}
\end{figure}
%
%
\section{The ideal observable}

In terms of the so--called Cartesian polarization observables,
the spin--dependent cross section can be written as\cite{meyer}
\begin{eqnarray}
\nonumber
\sigma (\xi, \vec P_b, \vec P_t, \vec P_f)
&=& \sigma_0(\xi)\left[1+\sum_i ((P_b)_iA_{i0}(\xi)+(P_f)_iD_{0i}(\xi))\right. \\
\nonumber
& &\phantom{\sigma_0(\xi)[1} \ +\sum_{ij}((P_b)_i(P_t)_jA_{ij}(\xi)+(P_b)_i(P_f)_jD_{ij}(\xi)) \\
& &\phantom{\sigma_0(\xi)[1}  \
+\left.\sum_{ijk}(P_b)_i(P_t)_j(P_f)_kA_{ij,k}(\xi) ... \right] \ .
\label{obsdef}
\end{eqnarray}
where $\sigma_0(\xi)$ is the unpolarized differential cross section, the
labels $i,j$ and $k$ can be either $x,y$ or $z$, and
$P_b$, $P_t$ and $P_f$ denote the polarization vector of beam, target and one
of the final state particles, respectively. All kinematic variables are
collected in $\xi$.

In
Refs.\cite{report,meyer,deepak} it was shown, that a measurement of 
the spin correlation parameters $A_{xx}, \ A_{yy}, \ A_{zz}$ as well
as the unpolarized  cross section allows to project on the individual initial
spin states. More precisely
\begin{eqnarray}
^1\sigma_0&=&\sigma_0(1-A_{xx}-A_{yy}-A_{zz}) \ ,\nonumber \\
^3\sigma_0&=&\sigma_0(1+A_{xx}+A_{yy}-A_{zz})
\  ,\nonumber \\
^3\sigma_1&=&\sigma_0(1+A_{zz}) \ ,
\label{spinwqs}
\end{eqnarray}
where the spin cross sections are labeled following the convention of Ref.
\cite{meyer} as $^{2S+1}\sigma_{M_S}$, with $S$ the total spin of the initial
state and $M_S$ its projection; $\sigma_0$ denotes the unpolarized cross
section.  Unfortunately, longitudinal polarization (needed for $A_{zz}$) is
not easy to prepare in a storage ring. However, the following linear
combination projects on spin triplet initial states and no longitudinal
polarization is needed\cite{unsers}:
\begin{eqnarray}
^3\sigma_\Sigma=\frac12(^3\sigma_0+^3\sigma_1)=\frac12\sigma_0(2+A_{xx}+A_{yy}) \ .
\end{eqnarray}
For $\vec p\vec p\to\Theta^+\Sigma^+$, only negative parity states contribute
to $^3\sigma_\Sigma$; thus only in case of a negative parity $\Theta^+$
$s$--waves are allowed in the final state. If we assume the $\Theta^+$ to be
an isoscalar, the reaction $\vec p\vec n\to \Theta^+ \Lambda$ gives the same
amount of information, however, positive and negative parity change their
roles.

It is well known that for large momentum transfer reactions in the near
threshold regime the energy dependence of a partial wave characterized by
angular momentum $l$ is given by $(p/\Lambda)^l$, where $\Lambda$ denotes the
intrinsic scale of the production process---for reactions of the type $NN\to
B_1B_2$ this is typically given by the momentum transfer
$$p\sim \sqrt{(M_{B_1}+M_{B_2}-2M_N)M_N} \ .$$
For an extensive discussion of
this type of reaction we refer to Ref.\cite{report}.  It is thus sufficient
to measure the energy dependence of $^3\sigma_\Sigma$ for small excess
energies to pin down the parity of the $\Theta^+$: in case of a positive
parity in the $pp$ channel it scales as phase--space times an odd polynomial
in the excess energy $Q=\sqrt{s}-\sqrt{s_0}$, where $s_0$ denotes the
threshold energy.  On the other hand, if the resonance has negative parity the
energy dependence should be that of phase--space times an even polynomial.  It
should be stressed that these considerations apply only for outgoing cms
momenta significantly smaller than $\Lambda$---for larger energies no general
statement on the energy dependence is possible in a model--independent way.
This is quite obvious once it is recalled that there are rigorous
energy--dependent bounds on the strength of the individual partial waves set
by unitarity---thus there should not be an unlimited growth.

The near threshold properties of $^3\sigma_\Sigma$ are shown in the lower line
of Fig. \ref{int}, where the results for different models (for more details
about these see next section and the appendix of Ref.\cite{unser2}) for the
unpolarized cross section $\sigma_0$, the spin correlation coefficient
$A_{xx}$ and $^3\sigma_\Sigma$ are shown.  Although the energy dependence of
$\sigma_0$ as well as $A_{xx}$ is vastly different, reflecting the different
admixture of partial waves in the different models, the figure clearly
illustrates that the energy dependence of $^3\sigma_\Sigma$ is an unambiguous signal for the
parity of the $\Theta^+$.

In Ref.\cite{unser2} we also investigate if the spin transfer coefficient
$D_{xx}$ can be used for a parity determination and we refer the interested
reader to this paper. The remaining space available for these proceedings will
now be used to discuss the reliability and features of model calculations for
reactions of the type $NN\to B_1B_2$.

\begin{figure}[ht]
\epsfxsize=2.1in
\centerline{\epsfbox{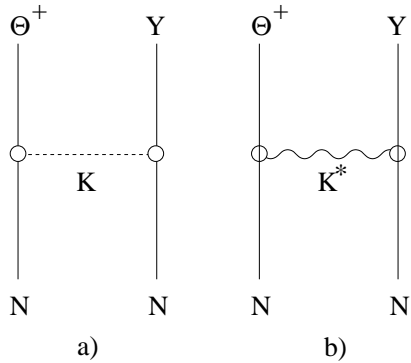}}   
\caption{\it Diagrams considered in the model calculations.  \label{dia}}
\end{figure}

\begin{figure}[ht]
\epsfxsize=4.1in
\centerline{\epsfbox{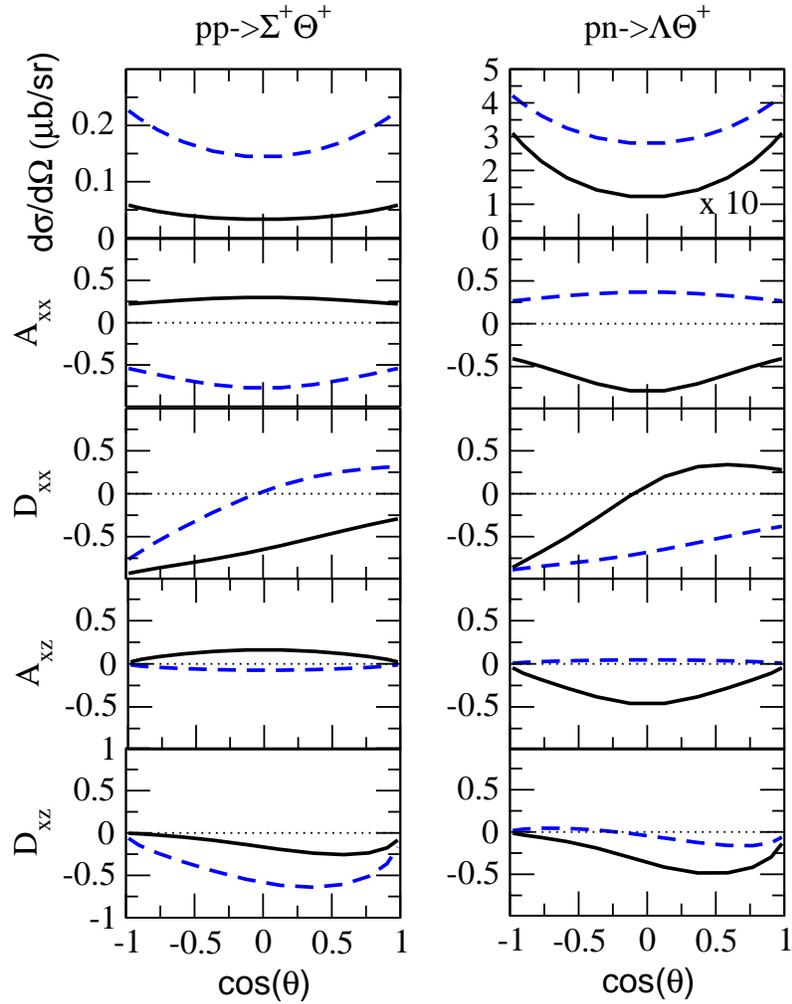}}   
\caption{\it Angular distributions for the $K$--exchange model calculated at
  an excess energy of 40 MeV. The meaning of the curves is as in 
Fig. \protect\ref{int}. The lower curve in the upper left panel
is scaled by a factor of 10 (as indicated in the panel). \label{kang}}
\end{figure}

\begin{figure}[ht]
  \epsfxsize=4.1in
\centerline{\epsfbox{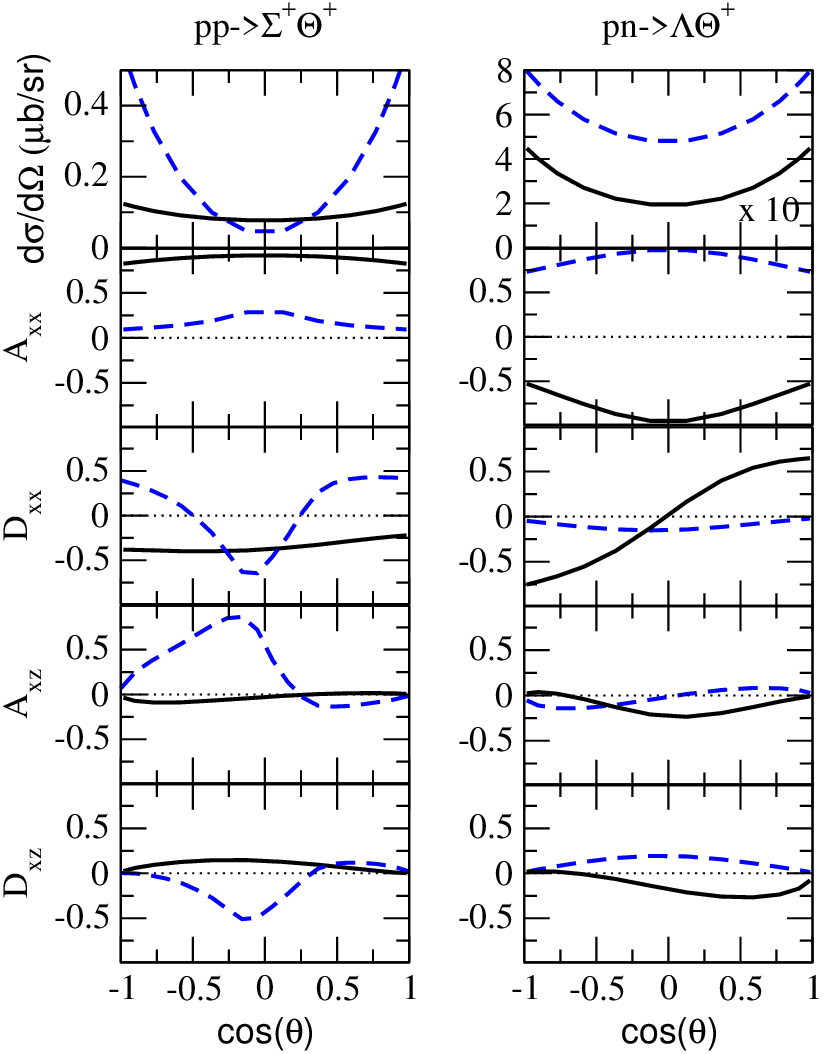}}
\caption{\it  Angular distributions for the $K+K^*$--exchange
 model calculated at
  an excess energy of 40 MeV.  The meaning of the curves is as in 
Fig. \protect\ref{int}. The lower curve in the upper left panel
is scaled by a factor of 10 (as indicated in the panel). \label{kkstang}}
\end{figure}

\section{Remarks on models}

As mentioned above the reactions $NN\to B_1B_2$ are characterized by a large
momentum transfer. This has two consequences: first of all the energy
dependence of the production process in the near threshold regime is fixed
model--independently; secondly, it is very difficult to construct a
reliable microscopic model for these reactions.

Probably the most clear illustration of the latter point is the fact that
there is not even a microscopic model available to describe the data on $pp\to
pp\pi^0$, although this is the first inelasticity of the $NN$ system and much
is known about all the subsystems. Only recently it was observed, that the
large momentum transfer as it occurs in inelastic $NN$ reactions leads to a
relative enhancement of pion loop contributions (see discussion in Ref.
\cite{report}). Given this it seems inappropriate to construct a model for the
reaction $NN\to \Theta^+Y$ that is quantitatively reliable, since here a lot
less is known about possible production mechanisms---there might even be quite
complicated production mechanisms of relevance, like the decay of a heavier
resonance as proposed in Ref.\cite{lipkin}.  In addition, it is well
established that the initial state interaction can have a significant effect
on observables. First of all it reduces the cross section typically by a
factor of 2--3 and secondly it introduces an additional phase to the
individual amplitudes; especially the latter effect can well change the shape
especially of polarization observables, for they are quite sensitive to the
relative phases of the contributing amplitudes.  Note, in reactions of the
type $NN\to \Theta^+ Y$, where the final state interaction is expected to be
weak, the relative phase of the amplitudes is largely introduced by the $NN$
interaction in the initial state. Through the Watson theorem\cite{gw} this
phase can be related to the $NN$ scattering phase--shifts. Although this is
true rigorously only in the elastic case, it is still reasonable to expect a
properly adjusted formula to also work for the $NN$ system at energies as high
as relevant for the $\Theta^+$ production\cite{withkanzo}. However, such a
detailed work is beyond the scope of this paper---especially, the initial
state interaction will not change significantly the energy dependence of
observables, as long as only energies close to the production threshold are
investigated.

As a result of all this as in Ref.\cite{unser2} we here take a more pragmatic
point of view: if all statements made above are indeed true
model--independently, they have to apply to any (realistic) model.  Since we
do not trust the overall scale of the model results and for a better
comparison of the results for the two different parities, the results for the
integrated observables in Fig. \ref{int} are normalized to 1 at 20 MeV.

We constructed a model where a scan through a wide range of parameters allowed
to study a large class of different effects. To be more specific, we included
the diagrams shown in Fig. \ref{dia}, fixing the coupling strength of the $K$
exchange and then varied the parameters for the $K^*$ exchange. The following
three models turned out to be representative for the many cases we studied,
namely a model with purely $K$ exchange, a model with purely $K^*$ exchange
and a mixture of both. In the third case the relative strength of the two
diagrams was adjusted such that for a positive parity $\Theta^+$ the $s$--wave
contributions of the two diagrams canceled in the $pp$ reaction\footnote{We
  also unsuccessfully tried to construct a model were the same happens for the
  negative parity.}. This
lead to a quite drastic energy dependence of $A_{xx}$.

It turned out that, although the energy dependence of the cross
section as well as $A_{xx}$ was very different for different models, the behavior of
$^3\sigma_\Sigma$ was always as described above.
 
One question repeatedly asked on the conference was that about the possible
production mechanisms. Given the problems mentioned above regarding the
construction of the production operator for $\Theta^+$ production in $NN$
collisions, it will most probably not be possible to unambiguously identify a
particular production mechanism as the most significant one from data on these
reactions directly.  However, what the polarization observables serve for is
to exclude particular production mechanisms. To be more concrete: as different
production mechanisms are typically characterized by different spin and
isospin dependencies, they will lead to quite different polarization
observables for the $pp$ and $pn$ induced reaction. This is illustrated in
Figs.~\ref{kang} and \ref{kkstang}, where the angular distributions of various
spin observables are shown for two different sets of model parameters (pure
$K$ exchange and $K+K^*$ exchange) for both channels.  Clearly, the two models
lead to very different angular dependencies of most of the observables shown.
Thus, these data could well be used to exclude particular production
mechanisms (once the initial state interaction is included as described above)
and thus improve our understanding of the hadronic interactions of the
$\Theta^+$ (... if it exits).

\section{Summary and outlook}

In summary we have argued that the most promising method to model
independently determine the parity of narrow resonances is a measurement of
$\vec N\vec N\to B_1 B_2$: the energy dependence of the spin triplet cross
section, given by $^3\sigma_\Sigma=\frac12\sigma_0(2+A_{xx}+A_{yy})$, is the
ideal observable. There is also a chance that $\sigma_0D_{xx}$ also allows to
determine the parity---for details on this we refer to Ref.\cite{unser2}.

 In
addition, especially the angular dependence of the large number of
existing polarization observables will put strong constraints on the allowed
production mechanisms.

In this project theory did its part---now we have to wait for the experimental
realization.  We want to close with a few comments on the prospects of a
measurement of $A_{xx}$ for $NN\to \Theta^+ Y$. So far a measurement for the
unpolarized cross section in the $pp$ channel is completed by the TOF
collaboration at the COSY accelerator\cite{tof}: with a statistical
significance of about 4.5 $\sigma$ a total cross section of 400 nb was
extracted from data on $pp\to \Sigma^+ pK^0$. At COSY polarized beams are
routinely available and a frozen spin target is currently being adopted to the
COSY conditions. The first double polarized measurement is expected for summer
2005.

{\bf Acknowledgments}

C.H. thanks the organizers for a very informative, inspiring, educating,
enjoyable, and perfectly organized workshop.

\end{document}